\documentclass[12pt]{article}
\usepackage{amssymb}
\def\beq{\begin{equation}}
\def\eeq{\end{equation}}
\usepackage[all,2cell,dvips]{xy}
\def\IR{\relax{\rm I\kern -.18em R}}

\begin{document}
\title{Infinite sequence of new conserved quantities for $N=1$ SKdV and the supersymmetric cohomology }
\author{ \Large S. Andrea*, A. Restuccia**, A. Sotomayor***}
\maketitle{\centerline {*Departamento de Matem\'{a}ticas,}}
\maketitle{\centerline{**Departamento de F\'{\i}sica}}
\maketitle{\centerline{Universidad Sim\'on Bol\'{\i}var}}
\maketitle{\centerline{***Departamento de Ciencias B\'{a}sicas}}
\maketitle{\centerline{Unexpo, Luis Caballero Mej\'{\i}as }}
\maketitle{\centerline{e-mail: sandrea@usb.ve, arestu@usb.ve,
sotomayo81@yahoo.es }}
\begin{abstract}An infinite sequence of new non-local conserved
quantities for $N=1$ Super KdV (SKdV) equation is obtained. The
sequence is constructed, via a Gardner trasformation, from a new
conserved quantity of the Super Gardner equation. The SUSY
generator defines a nilpotent operation from the space of all
conserved quantities into itself. On the ring of
$C_\downarrow^\infty$ superfields the local conserved quantities
are closed but not exact. However on the ring of
$C_{NL,1}^\infty$ superfields, an extension of the
$C_\downarrow^\infty$ ring, they become exact and equal to the
SUSY transformed of the subset of odd non-local conserved
quantities of the appropriate weight. The remaining odd non-local
ones generate closed geometrical objects which become exact when
the ring is extended to the $C_{NL,2}^\infty$ superfields and
equal to the SUSY transformed of the new even non-local conserved
quantities we have obtained. These ones fit exactly in the SUSY
cohomology of the already known conserved quantities.
\end{abstract}

\section{Introduction}Supersymmetric integrable systems are an
interesting scenario to analyze ADS/CFT correspondence
\cite{Maldacena}, in particular in relation to $N=4$ Super
Yang-Mills models. Super conformal algebras are also realized in
Super KdV equations(SKdV) \cite{Mathieu,Mathieu1,Manin}. SKdV
equations are also directly related to supersymmetric quantum
mechanics. In fact, the whole SKdV hierarchy arises from the
asymptotic expansion of the Green's function of the Super heat
operator \cite{Andrea2}.

One of the main properties of the integrable systems is the
presence of an infinite sequence of conserved quantities.

For SUSY integrable systems the complete structure of conserved
quantities has not been understood.

For $N=1$ SKdV an infinite sequence of local conserved quantities
was found in \cite{Mathieu}. It was then observed, by analyzing
the symmetries of SKdV, the existence of odd non-local conserved
quantities \cite{Kersten1,Kersten2}. In \cite{Dargis} they were
obtained from a Lax formulation of the Super KdV hierarchy and
generated from the super residue of a fractional power of the Lax
operator.

In \cite{Andrea3} all these odd non-local conserved quantities
were obtained from a single conserved quantity of the SUSY Gardner
equation(SG), which was introduced in \cite{Mathieu}, see also
\cite{Andrea}. So far only two conserved quantities of the SG
equation are known, the one generating all the local conserved
quantities of SKdV, it is of even parity and of dimension 1, and
the above mentioned one of odd parity and of dimension
$\frac{1}{2}$.

In this paper we introduce a new infinite sequence of non-local
conserved quantities of $N=1$ SKdV. We construct them via a
Gardner map \cite{Gardner}, from a new conserved quantity of SG.
It is non-local and it has even parity and dimension 1. We then
introduce the SUSY cohomology in the space of conserved
quantities and obtain the relation between all the conserved
quantities local, odd non-local and even non-local. In this sense
it is natural the introduction of the new even conserved
quantities. The SUSY cohomology we introduce here is used for the
$N=1$ KdV system, however several of the arguments are general,
beyond SKdV, and we expect they will be useful in the analysis of
other integrable systems.

In section 2 we present basic definitions, in sections 3 and 4
the new conserved quantities and in section 5 the SUSY cohomology.

\section{Basic facts}
A first step in order to analize $N=1$ SKdV and its known local
and odd non-local conserved quantities is to consider the ring of
polynomials with one odd generator $a_1$ and a superderivation
$D$ defined by $Da_n=a_{n+1}$ ($n\in \mathbb{N}$). The elements
$a_n$ have the same parity as the positive integer $n$ and
satisfy $a_{n_1}a_{n_2}=\pm a_{n_2}a_{n_1}$ with a minus sign
only in the case when $n_1$ and $n_2$ are odd. On the products
$D$ acts following the rule
$D(a_{n_1}a_{n_2})=(Da_{n_1})a_{n_2}+{(-1)}^{n_1}a_{n_1}Da_{n_2}$.
The explicit algebraic presentation is given by the ring
$\mathcal{A}$ of elements of the form $b(a_1,a_2,\ldots)$, with
$b$ any polynomial, and $D=a_{2}\frac{\partial }{\partial
a_{1}}+a_{3}\frac{\partial }{\partial a_{2}}+\cdots$

A second step to complete the construction is to extend the ring $
\mathcal{A}$ with a new set of generators of the form
$\lambda_n=D^{-1}h_n$, with $h_n$ being the non trivial integrands
of the known even local conserved quantities. An element of the
new ring $\widetilde{\mathcal{A}}$ is a polynomial of the form $
\widetilde {b}( a_1,a_2,\ldots,\lambda_1,\lambda_3,\ldots)$ and
the corresponding superderivation is given by $
\widetilde{D}=D+P$ with $P=h_1\frac{\partial }{\partial
\lambda_{1}}+h_3\frac{\partial }{\partial
\lambda_{3}}+h_5\frac{\partial }{\partial \lambda_{5}}+\cdots$

 This algebraic presentation is directly connected to
the analytical one by the substitution $a_1\leftrightarrow \Phi$,
where $\Phi$ is a superfield, that is, a function $
\Phi:\mathbb{R}\rightarrow \Lambda$, with $\Lambda$ a Grassmann
algebra one of whose generators $\theta$ has been singled out. In
this setting $D=\frac{\partial}{\partial \theta}+\theta
\frac{\partial}{\partial x}$ and $ \Phi(x,\theta)=\xi(x)+\theta
u(x),$ with $t$ an implicit variable. The parity of the generator
$a_1$ requires $\xi$ to be odd and $u$ to be even for each $x\in
\mathbb{R}$.

The covariant derivative $D$ has the property
$D^2=\frac{\partial}{\partial x}$ acting on differentiable
superfields, and the generator of SUSY transformations
$Q=-\frac{\partial}{\partial \theta}+\theta
\frac{\partial}{\partial x}$ satisfies $DQ+QD=0.$

In order to realize ``integration" the ring of infinitely
differentiable superfields $C^\infty( \mathbb{R},\Lambda)$ must
be restricted. We then introduce the ring of Schwartz
superfields, that is,
\[C^\infty_\downarrow(\mathbb{R},\Lambda)=\left\{\Phi \in
C^\infty( \mathbb{R},\Lambda)/\lim_{x\rightarrow \pm
\infty}x^p\frac{\partial^q}{\partial x^q}\Phi=0\right\}, \] for
every $p,q\geq 0$. The ``superintegration" in this space is well
defined and we write $\int \Phi dxd\theta=\int_{-\infty}^\infty
udx.$ It is natural to introduce the space of integrable
superfields $C^\infty_I(\mathbb{R},\Lambda)=\left\{\Phi \in
C^\infty(\mathbb{R},\Lambda)/\frac{\partial}{\partial
\theta}\Phi\in C^\infty_\downarrow(\mathbb{R},\Lambda)\right\}$
and clearly $C^\infty_\downarrow(\mathbb{R},\Lambda)\subset
C^\infty_I(\mathbb{R},\Lambda)$.

The next ring to be  considered allows us to deal with the known
odd non-local conserved quantities. We introduce then the ring of
non-local superfields
$C^\infty_{NL,1}(\mathbb{R},\Lambda)=\left\{\Phi \in C^\infty(
\mathbb{R},\Lambda)/D\Phi\in
C^\infty_\downarrow(\mathbb{R},\Lambda) \right\}$. We have
$C^\infty_\downarrow(\mathbb{R},\Lambda)\subset
C^\infty_{NL,1}(\mathbb{R},\Lambda) \subset
C^\infty_I(\mathbb{R},\Lambda)$. If $\Psi=U+\theta V \in
C^\infty_{NL,1}(\mathbb{R},\Lambda) $, we take, as in the case of
integrable superfields $\int \Psi dxd\theta=\int_{-\infty}^\infty
Vdx.$ In conclusion, for a given superfield to be integrable it
is sufficient that the application of $\frac{\partial}{\partial
\theta}=D|_{\theta=0}$ to it gives a Schwartz superfield.

The crucial fact we use in the formulation for next sections is
that $C^\infty_\downarrow(\mathbb{R},\Lambda)$ is an ideal of
$C^\infty_{NL,1}(\mathbb{R},\Lambda)$.

We remind that a candidate for a conserved quantity (for example
in the $C_\downarrow^\infty(\mathbb{R},\Lambda)$ ring) associated
to a given partial differential equation (PDE)
\beq\Phi_t=k(\Phi,D\Phi,D^2\Phi,\ldots)\label{pde}\eeq may be
presented by $H=\int h(\Phi,D\Phi,D^2\Phi,\ldots)dxd\theta$
($k,h\in C_\downarrow^\infty(\mathbb{R},\Lambda) $). Then, $H$ is
a conserved quantity if $H_t=0$, which is equivalent to have
$\frac{d}{ds}|_{s=0}H(\Phi+s\Psi)=0$ whenever
$\Psi(x,t)=\frac{d}{dt}\Phi(x,t)=k(\Phi(x,t),D\Phi(x,t),\ldots)$.
 In terms of functional derivatives a necessary and sufficient condition for $H$ to be a conserved quantity of
 (\ref{pde}) is to have $\delta
 h=Dg$ for some $g\in C^\infty_\downarrow(\mathbb{R},\Lambda)$. This condition is used in section 4.

 To describe the known local conserved quantities the last setting is
 sufficient and the ring $ \mathcal{\widetilde{A}}$ may be reduced by
 considering only polynomials of the form $b(a_1,a_2,\ldots)$.

 For the study of the known
odd non-local conserved quantities we must consider the complete
ring $ \mathcal{\widetilde{A}}$. For example, the fourth odd
non-local known conserved quantity is given by
\beq\int\left\{\frac{1}{24}{(D^{-1}\Phi)}^4-\frac{1}{2}{(D\Phi)}^2+(D^{-1}\Phi)D^{-1}(\Phi
  D\Phi)\right\}dxd\theta\label{fnlcq},\eeq and the terms $D^{-1}\Phi$ and $D^{-1}(\Phi
  D\Phi) $ clearly belongs to the ring extension $C_{NL,1}^\infty$ mentioned
  before, where $\Phi=h_1$ and $\Phi D\Phi=h_3$. We put $a_0\equiv
  \lambda_1=D^{-1}\Phi$ in accordance with notation of \cite{Andrea3}.
  The formal analysis of this type of conserved quantities is
  similar to the local ones but in this case the ring $C^\infty_{NL,1}(\mathbb{R},\Lambda)$ plays a fundamental
  role.

  To obtain the known local and non-local conserved quantities we
  start with the pair of SuperKdV and Super Gardner
  equations given by
  \beq
\Phi_t=D^6\Phi+3D^2(\Phi D\Phi),\label{susyeq}\eeq and \beq
\chi_t=
 D^6\chi+3D^2(\chi D\chi)-3\epsilon^2
 (D\chi)D^2(\chi D\chi),\label{superg}\eeq with $\epsilon$ a
 given parameter and $\chi=\sigma +\theta w$ another odd superfield \cite{Mathieu}.

 (\ref{susyeq}) and (\ref{superg}) are connected by the Super Gardner
 map given by \beq \Phi=\chi+\epsilon D^2\chi-\epsilon^2\chi D
 \chi.\label{strag}\eeq It holds \beq
\begin{array}{l}\Phi_t-D^6\Phi-3D^2\left(\Phi D\Phi
\right)=\left[1+\epsilon D^2-\epsilon^2\left(D\chi+\chi
D \right) \right]\cdot \\
\left\{\chi_t-D^6\chi-3D^2\left(\chi D\chi
\right)+3\epsilon^2\left(D\chi \right)D^2\left(\chi D\chi \right)
\right\},\end{array}\label{factsupercampos}\eeq whenever $\Phi$
and $\chi$ are related by (\ref{strag}). We recall that
(\ref{strag}) maps solutions of (\ref{superg}) into
(\ref{susyeq}). The inverse is also true if we restrict the space
of possible solutions of (\ref{superg}) to formal series
$\chi=\sum_{n=0}^\infty a_n[\Phi]\epsilon^n$. $\int\chi dxd\theta$
is a conserved quantity of (\ref{superg}), it induces the infinite
known local conserved quantities for (\ref{susyeq}). In the
language of \cite{Andrea3} the right members of
(\ref{susyeq}),(\ref{superg}),(\ref{strag}) are denoted by
$g,f,r$ respectively, with $g\in \mathcal{A}$ and $f,r\in
\mathcal{A}[\epsilon]$. Condition (\ref{factsupercampos}) is
equivalent to $g\circ r=r^\prime f$ and is necessary and
sufficient for (\ref{strag}) to map solutions of Super Gardner to
SuperKdV.

In \cite{Andrea3} it was shown that $\int \exp^{(\epsilon
D^{-1}\chi)}dxd\theta$ is also a conserved quantity for
(\ref{superg}). This induces the known infinite sequence of odd
non-local conserved quantities of SKdV.

\section{Infinite sequence of new non-local conserved quantities of SKdV}
We will introduce in this section a new non-local conserved
quantity of the Super Gardner equation. From it, following the
previous section, we may obtain an infinite set of new non-local
conserved quantities of SKdV. There are two already known
conserved quantities of the Super Gardner equation. The first one
\cite{Mathieu} provides the infinite set of local conserved
quantities of SKdV equation. The other one \cite{Andrea3} give
rise to the infinite set of odd non-local conserved quantities of
SKdV, originally found in \cite{Kersten1} and also obtained from
the Lax operator in \cite{Dargis}. This two conserved quantities
of Super Gardner equation exhausts all the known conserved
quantities of SKdV equation. We will now introduce a new infinite
set of even non-local conserved quantities of SKdV.

The quantity $H_G$, \beq \begin{array}{ll}
H_G=\int\left\{D^{-1}\left[\frac{\exp\left(\epsilon
D^{-1}\chi\right)+\exp\left(-\epsilon D^{-1}\chi
\right)-2}{2\epsilon^2} \right]\right.+  \\ \left.
+\frac{1}{2}\left[\frac{\exp\left(\epsilon D^{-1}\chi
\right)-1}{\epsilon}\right] D^{-1}\left[\frac{\exp\left(-\epsilon
D^{-1}\chi\right)-1}{\epsilon} \right]
\right\}dxd\theta\end{array}\label{consgard1}\eeq where $\chi\in
C^\infty_\downarrow$, exists and is conserved by every solution
of the Super Gardner equation. It has the same parity as the
local conserved quantities and opposite to the already known
non-local conserved quantities of SKdV. When we apply the inverse
Gardner transformation we obtain an infinite set of well defined
even non-local conserved quantities of SKdV.

The terms \[D^{-1}\left[\frac{\exp\left(\epsilon
D^{-1}\chi\right)+\exp\left(-\epsilon D^{-1}\chi
\right)-2}{2\epsilon^2} \right]\] and
\[D^{-1}\left[\frac{\exp\left(-\epsilon
D^{-1}\chi\right)-1}{\epsilon} \right]\] do not belong to
$C_{NL,1}^\infty$, but to $C_{NL,2}^\infty=\left\{\Phi \in
C^\infty( \mathbb{R},\Lambda)/D^2\Phi\in
C^\infty_\downarrow(\mathbb{R},\Lambda) \right\}$.

Although each term is not integrable the complete integrand
belongs to $C_I^\infty$.

We notice that (\ref{consgard1}) may be rewritten as \beq
H_G=\int\frac{1}{2}D^{-1}\left\{D\left[\frac{\exp\left(\epsilon
D^{-1}\chi \right)-1}{\epsilon}\right]
D^{-1}\left[\frac{\exp\left(-\epsilon
D^{-1}\chi\right)-1}{\epsilon}\right]\right\}dxd\theta.\label{consgard2}\eeq
We may then perform the $\theta$ integration and obtain \beq
H_G=\int_{-\infty}^\infty\frac{1}{2}\left\{D\left[\frac{\exp\left(\epsilon
D^{-1}\chi \right)-1}{\epsilon}\right]
D^{-1}\left[\frac{\exp\left(-\epsilon
D^{-1}\chi\right)-1}{\epsilon}\right]\right\}dx\label{consgard3}\eeq
which under the assumption $\chi\in C_\downarrow^\infty$ is a
well defined integral. The first two new conserved quantities of
Super KdV which may be obtained from the inverse Gardner
transformation are \beq H^{NL}_1=\int\frac{1}{2}D^{-1}(\Phi
D^{-2}\Phi)dxd\theta, \label{ncq1} \eeq \beq\begin{array}{ll}
H^{NL}_3=\int\left[-\frac{1}{2}D^{-1}(\Phi
D^2\Phi)+D^{-1}(D\Phi\cdot \Phi \cdot
D^{-2}\Phi)+\frac{1}{24}D^{-1}{(D^{-1}\Phi)}^4-\right. \\\left.
-\frac{1}{6}D^{-2}\Phi{(D^{-1}\Phi)}^3+\frac{1}{8}{(D^{-1}\Phi)}^2D^{-1}{(D^{-1}\Phi)}^2\right]dxd\theta\end{array}
\label{ncq2}\eeq which exactly agree with the two non-local
conserved quantities obtained in \cite{Andrea1} by the
supersymmetric recursive gradient procedure. The proofs of
existence in \cite{Andrea1} for the recursive gradient procedure
were only given for local conserved quantities, there are no
proofs in the literature for the recursive gradient procedure
involving non-local quantities. Hence the existence of the
infinite set of conserved quantities is not guaranteed from that
approach.

(\ref{ncq2}) may be rewritten using (\ref{consgard2}) as \beq
H_3^{NL}=\int D^{-1}\left[\Phi\left(-\frac{1}{2}D^2\Phi-D\Phi\cdot
D^{-2}\Phi-\frac{1}{2}D^{-2}\Phi\cdot
{\left(D^{-1}\Phi\right)}^2+\frac{1}{4}D^{-1}\Phi\cdot
D^{-1}{\left(D^{-1}\Phi\right)}^2 \right)\right]dxd\theta,
\label{ncqsteve}\eeq where the integrand is manifestly in
$C_I^\infty.$

In the next section we prove the claims of existence and
conservation of $H_G$ and consequently of the new set of infinite
conserved quantities of SKdV.

\section{Conservation of $H_G$ under the Super Gardner flow}
The proposed non-local quantity $H_G$ for Super Gardner, written
in terms of an appropriate analytical extension of the ring $
\mathcal{A}$ (we make use of an abuse of notation taking
$D^{-1}\chi\leftrightarrow a_0,\chi\leftrightarrow a_1$ an so on),
is defined by \beq h_G=\left(\exp^{\epsilon
a_0}-1\right)D^{-1}\left(\exp^{-\epsilon
a_0}-1\right)+D^{-1}\left(\exp^{\epsilon a_0}+\exp^{-\epsilon
a_0}-2\right). \label{f6}\eeq For $h_G$ to be integrable, $h_G\in
C_I^\infty,$ it must satisfy, $\frac{\partial}{\partial
\theta}h_G\in C_\downarrow^\infty.$ From $a_1\leftrightarrow \chi=
\sigma+\theta w$ and
$D^{-1}a_1=a_0=\int_{-\infty}^xw+\theta\sigma$ it follows that
$e^{\epsilon a_0}-1\in C_I^\infty.$ The inclusion
\[D^{-1}\frac{\partial}{\partial \theta}C_I^\infty\subset
C_\downarrow^\infty\] gives $D^{-1}\frac{\partial}{\partial
\theta}\left(e^{\epsilon a_0}-1 \right)\in C_\downarrow^\infty.$

Using $D^{-1}\frac{\partial}{\partial
\theta}+\frac{\partial}{\partial \theta}D^{-1}=
\mathrm{\:the\:}\mathrm{\:identity\:}\mathrm{\:operator\:},$ we
compute
\begin{eqnarray*} \frac{\partial h_G}{\partial \theta} &=& C_\downarrow^\infty
+\left(e^{\epsilon a_0}-1
\right)\left(I-D^{-1}\frac{\partial}{\partial \theta}
\right)\left(e^{-\epsilon a_0}-1
\right)+\left(I-D^{-1}\frac{\partial}{\partial \theta}
\right)\left(e^{\epsilon a_0}+e^{\epsilon a_0}-2 \right) \\
&=& C_\downarrow^\infty+\left(e^{\epsilon a_0}-1
\right)\left(e^{-\epsilon a_0}-1 \right)+\left(e^{\epsilon
a_0}+e^{-\epsilon a_0}-2 \right) \\ &=& C_\downarrow^\infty.
\end{eqnarray*} This proves that the proposed non-local conserved
quantity is integrable.

Next it must be shown that when $a_1$ is replaced by $a_1+sf$ and
$a_0$ by $a_0+sg$ we have
\[\frac{d}{ds}|^{s=0}h_G\left(a_0+sg,a_1+sf,a_2+sDf,\ldots   \right)\in DC_\downarrow^\infty. \]
Using $\delta \Omega(a_0,a_1,\ldots)$ to denote
$\frac{d}{ds}|^{s=0}\Omega\left(a_0+sg,a_1+sf,a_2+sDf,\ldots
\right)$ for any $\Omega(a_0,a_1,\ldots),$ we have
\[\frac{1}{\epsilon}\delta e^{\epsilon a_0}=\left(e^{\epsilon a_0} \right)g.   \]
Thus \begin{eqnarray*} \frac{1}{\epsilon}\delta h_G&=&e^{\epsilon
a_0}gD^{-1}\left(e^{-\epsilon a_0}-1 \right)+\left(e^{\epsilon
a_0}-1
\right)D^{-1}\left(-e^{-\epsilon a_0}g \right) \\
&+&D^{-1}\left(\left(e^{\epsilon a_0}-e^{-\epsilon a_0} \right)g
\right).
\end{eqnarray*} However $D(e^{\epsilon a_0})(F_0+\epsilon
F_1+\epsilon^2F_2)=e^{\epsilon a_0}g.$

Therefore, with $F(\epsilon)=F_0+\epsilon F_1+\epsilon^2F_2,$
\begin{eqnarray*}\frac{1}{\epsilon}\delta h_G &=& \left(De^{\epsilon a_0}F(\epsilon) \right)
D^{-1}\left(e^{-\epsilon a_0}-1 \right)\left(e^{\epsilon a_0}-1
\right)e^{-\epsilon a_0}F(-\epsilon) \\ &+& e^{\epsilon
a_0}F(\epsilon)-e^{-\epsilon a_0}F(-\epsilon).
\end{eqnarray*}
At this point it is crucial that $F(\epsilon)\in
C_\downarrow^\infty,$ as is $e^{\epsilon
a_0}F(\epsilon)D^{-1}\left(e^{-\epsilon a_0}-1 \right).$
Therefore, except for a term in $DC_\downarrow^\infty$,

\begin{eqnarray*} \frac{1}{\epsilon}\delta h_G &=& e^{\epsilon a_0}F(\epsilon)\left(
e^{-\epsilon a_0}-1\right)-\left(1-e^{-\epsilon a_0}
\right)F(-\epsilon) \\ &+& e^{\epsilon
a_0}F(\epsilon)-e^{-\epsilon a_0}F(-\epsilon)\end{eqnarray*}
\begin{eqnarray*}\frac{1}{\epsilon}\delta h_G &=& F(\epsilon)-e^{\epsilon a_0}F(\epsilon)-
F(-\epsilon)+e^{-\epsilon a_0}F(-\epsilon) \\ &+& e^{\epsilon
a_0}F(\epsilon)-e^{-\epsilon a_0}F(-\epsilon).
\end{eqnarray*}
Having arrived at \begin{eqnarray*}\frac{1}{\epsilon}\delta
h_G=F(\epsilon)-F(-\epsilon)=2\epsilon F_1,
\end{eqnarray*}
with $F_1=a_1a_4-a_2a_3=-D(a_1a_3).$

This verifies that $\delta h_G\in DC_\downarrow^\infty$, showing
in last term that \beq\int
h_Gdxd\theta=\int{\left\{\left(e^{\epsilon a_0}-1
\right)D^{-1}\left(e^{-\epsilon a_0}-1
\right)+D^{-1}\left(e^{\epsilon a_0}+e^{-\epsilon a_0}-2
\right)\right\}}_{a_0=D^{-1}\chi}dxd\theta\label{f7}\eeq is indeed
a non-local conserved quantity for the Super Gardner equation
\[\frac{\partial}{\partial
t}\chi(x,t)=f(\chi,D\chi,D^2\chi,\ldots),
\] with $f$ having the particular expression given by
\[f=(a_7+3a_1a_4+3a_2a_3)-3\epsilon^2(a_1a_2a_4+a_2^2a_3).  \]

\section{The SUSY cohomology on the space of conserved quantities}
The invariance under supersymmetry of SKdV equations implies that
the SUSY transformations of conserved quantities are also
conserved quantities. That is, if $H=\int h,h\in C_I^\infty,$ is
conserved under the SKdV flow then
\[\delta_QH:=\int Qh\] is also a conserved quantity.

The operation $\delta_Q$ acting on functionals of the above form
is well defined since under the change, leaving $H$ invariant,
\[h\rightarrow h+Dg\] with $g\in C_\downarrow^\infty$, we have
\[Qh\rightarrow Qh+QDg=Qh+D(-Qg)\] where $Qg\in
C_\downarrow^\infty$.

$\delta_Q$ is a superderivation satisfying $\delta_Q\delta_Q=0.$
In fact, \[\delta_Q\delta_QH=\int Q^2h=-\partial_\theta
h|_{-\infty}^\infty=0\] since $h\in C_I^\infty.$

For the local conserved quantities of SKdV, which we denote
$H_{2n+1}(\Phi),n=0,1,\ldots,$ we have
\beq\delta_QH_{2n+1}(\Phi)=0,n=0,1,\ldots,\label{cohom0}\eeq
where the index $2n+1$ denotes the dimension of $H_{2n+1}.$

If we consider the ring $C_\downarrow^\infty$ of superfields,
$H_{2n+1}$ is closed but not exact. However if we extend the ring
to the superfields $C_{NL,1}^\infty$, $C_\downarrow^\infty\subset
C_{NL,1}^\infty,$ then $H_{2n+1}$ becomes exact and it is
expressed in terms of $\delta_QH_{2n+\frac{1}{2}},n=0,1,\ldots$
where $H_{n+\frac{1}{2}},n=0,1,\ldots$ denote the odd non-local
conserved quantities of SKdV \cite{Kersten1,Dargis,Andrea3}, they
have dimension $n+\frac{1}{2}$. The remaining
$H_{2n+\frac{3}{2}}^{NL},n=0,1,\ldots$ plus a polynomial of lower
dimensional conserved quantities is closed but not exact in
$C_{NL,1}^\infty$, however if we extend the ring of superfields
to $C_{NL,2}^\infty$ they become exact and equal to
$\delta_QH_{2n+1}^{NL}$, where $H_{2n+1}^{NL}$ are the even
non-local conserved quantities we have introduced in the previous
section. They have dimension $2n+1$. To obtain the exact relation
between them we use the conserved quantities of the Super Gardner
equation.

We denote them $G_1,G_{\frac{1}{2}}^{NL}$ and $G_1^{NL}.$ We have
\beq G_1=\int
\chi=\sum_{n=0}\epsilon^{2n}H_{2n+1}\label{cohom1}\eeq \beq
G_{\frac{1}{2}}^{NL}=\int \frac{\exp(\epsilon
D^{-1}\chi)-1}{\epsilon}=\sum_{n=0}\epsilon^nH_{n+\frac{1}{2}}^{NL}
\label{cohom2}\eeq \beq  \begin{array}{ll} H_G\equiv
G_1^{NL}=\int\left\{D^{-1}\left[\frac{\exp\left(\epsilon
D^{-1}\chi\right)+\exp\left(-\epsilon D^{-1}\chi
\right)-2}{2\epsilon^2} \right]\right.+  \\ \left.
+\frac{1}{2}\left[\frac{\exp\left(\epsilon D^{-1}\chi
\right)-1}{\epsilon}\right] D^{-1}\left[\frac{\exp\left(-\epsilon
D^{-1}\chi\right)-1}{\epsilon} \right]
\right\}=\sum_{n=0}\epsilon^{n}H_{n+1}^{NL}\end{array}
\label{cohom3}\eeq where $\chi \in C_\downarrow^\infty$.

The odd powers of $\epsilon$, in(\ref{cohom3}), do not provide new
conserved quantities of SKdV. For example \beq
H_2^{NL}=\frac{1}{2}H_{\frac{1}{2}}^{NL}H_{\frac{3}{2}}^{NL}.
\label{secquanttriv}\eeq

We then have \[\delta_QG_1=\int
Q\chi=\chi|_{-\infty}^{+\infty}=0\] hence we obtain
(\ref{cohom0}).

We also have \beq
\begin{array}{cc}\delta_QG_{\frac{1}{2}}^{NL}=\frac{\exp(\epsilon G_1)-1}{\epsilon}=G_1+\frac{1}{2}\epsilon G_1^2+\cdots
\\ =\sum_{n=0}\epsilon^{2n}H_{2n+1}+\frac{1}{2}\epsilon{\left(\sum_{n=0}\epsilon^{2n}H_{2n+1}\right)}^2+\cdots\end{array}\label{cohom6} \eeq
from which we obtain the relation between the odd non-local and
local conserved quantities. In particular we get \beq
\delta_QH_\frac{1}{2}^{NL}=H_1, \label{cohom4}\eeq and
\[\delta_QH_\frac{3}{2}^{NL}=\frac{1}{2}H_1^2=\delta_Q(\frac{1}{2}H_1H_\frac{1}{2}^{NL})\]
that is
\beq\delta_Q\left(H_\frac{3}{2}^{NL}-\frac{1}{2}H_1H_\frac{1}{2}^{NL}
\right)=0.\label{cohom5}\eeq This is the generic situation , from
(\ref{cohom6}), $H_{2n+1},n=0,1,\ldots$ is expressed as an exact
quantity in terms of $\delta_Q$[$H_{2n+\frac{1}{2}}^{NL}+\Sigma$
products of lower dimensional conserved quantities] while
[$H_{2n+\frac{3}{2}}+\Sigma$ products of lower dimensional
conserved quantities] is closed in the ring $C_{NL,1}^\infty.$ If
we extend the ring of superfields to $C_{NL,2}^\infty$, then the
closed quantity becomes exact and expressed in terms of
$H_{2n+1}^{NL},n=0,1,\ldots $ The integrand of $H_{2n+1}^{NL}$ is
expressed in terms of superfields in $C_{NL,2}^\infty$ with the
property that the whole integrand belongs to $C_I^\infty.$ In the
case of $H_{n+\frac{1}{2}}^{NL}$ the integrand is expressed in
terms of superfields in $C_{NL,1}^\infty\subset C_I^\infty,$
hence each term is integrable. For example, $H_1^{NL}$ (see
(\ref{ncq1})) may be expressed as \beq
H_1^{NL}=\int\left[D^{-1}\left(\frac{1}{2}{\left(D^{-1}\Phi\right)}^2\right)-\frac{1}{2}D^{-1}\Phi
\cdot D^{-1}\left(D^{-1}\Phi\right)\right]\label{cohom7}\eeq each
term in the integrand belongs to $C_{NL,2}^\infty$, it is not
integrable but the combination is in $C_I^\infty$. This
expression is in terms of $D^{-1}h$ where $h$ are the integrands
of previously known conserved quantities
$H_{n+\frac{1}{2}}^{NL},H_{2n+1}$. In this particular case
\[H_{\frac{3}{2}}^{NL}=\int \frac{1}{2}{(D^{-1}\Phi)}^2,\]
\[H_1=\int \Phi,\] \[H_{\frac{1}{2}}^{NL}=\int D^{-1}\Phi.\]
This is also a generic property of $H_{2n+1}^{NL},n=0,1,\ldots$
and as we already knew of $H_{n+\frac{1}{2}}^{NL},n=0,1,\ldots$
whose integrands may be expressed in terms of polynomials in
$D^{-1}h$ where $h$ are the integrands of the local conserved
quantities $H_{2n+1}$.

From (\ref{cohom7}) we have
\[\delta_QH_1^{NL}=H_{\frac{3}{2}}^{NL}-\frac{1}{2}H_1H_{\frac{1}{2}}^{NL}\]
that is, the closed quantity becomes exact in $C_{NL,2}^\infty$.
Similar relations are obtained from (\ref{cohom3}) for higher
dimensional conserved quantities. The general formula is
\[\sum_{n=0}\epsilon^{n}\delta_QH_{n+1}^{NL}=\sum_{n=0}\epsilon^{2n}H_{2n+\frac{3}{2}}^{NL}-\frac{1}{2\epsilon}
\left[\exp\left(\sum_{n=0}\epsilon^{2n+1}H_{2n+1}\right)-1\right]
\left[\sum_{n=0}{(-\epsilon)}^{n}H_{n+\frac{1}{2}}\right]\]

We then have the following relations between the conserved
quantities of SKdV equation:

 \beq\xymatrix {H_1  &  &  H_3 &  & H_5  &  &H_7 &\cdots &
 \\ H_{\frac{1}{2}}^{NL}\ar[u]
   &H_{\frac{3}{2}}^{NL}  &  H_{\frac{5}{2}}^{NL}\ar[u] &H_{\frac{7}{2}}^{NL} & H_{\frac{9}{2}}^{NL}\ar[u] &\ldots
 &  &  &
 \\ &  H_1^{NL}\ar[u] &  & H_3^{NL}\ar[u] &  & H_5^{NL}\ar[u] & & } \label{diagram}\eeq
 where the arrow denotes the action of $\delta_Q$, up to lower
 dimensional conserved quantities as explained previously.

 The new conserved quantities $H_1^{NL},H_3^{NL},\ldots$ fit then
 exactly in the SUSY cohomology of the previously known conserved
 quantities.

\section{Conclusions}We found a new infinite sequence of non-local
conserved quantities of $N=1$ SKdV equations. They have even
parity and dimension $2n+1,n=0,1,\ldots$. We introduced the SUSY
cohomology in the space of conserved quantities: local, odd
non-local and even non-local. We found all the cohomological
relations between them.

Although we consider $N=1$ SKdV, we expect the SUSY cohomological
arguments to be valid in general for SUSY integrable systems.

\end{document}